\documentclass[12pt,preprint]{aastex}
\usepackage{epsfig}
\usepackage{amssymb}
\usepackage{graphicx}

\shortauthors{L. Cieza}

\begin{document}

\title{Imaging the water snow-line during a protostellar  outburst}

\author{
Lucas A. Cieza\altaffilmark{1,2}, Simon Casassus\altaffilmark{2,3}, John Tobin\altaffilmark{4},  Steven Bos\altaffilmark{4},  Jonathan P. Williams\altaffilmark{5}, Sebastian Perez\altaffilmark{2,3}, 
Zhaohuan Zhu\altaffilmark{6},
Claudio Caceres\altaffilmark{2,7},  Hector Canovas\altaffilmark{2,7}, Michael M. Dunham\altaffilmark{8},  Antonio Hales\altaffilmark{9},  Jose L. Prieto\altaffilmark{1}, David A. Principe\altaffilmark{1,2},  Matthias R. Schreiber\altaffilmark{2,7}, Dary Ruiz-Rodriguez\altaffilmark{10}, Alice Zurlo\altaffilmark{1,2,3}
}

\altaffiltext{1}{N\'ucleo de Astronom\'ia, Facultad de Ingenier\'ia, Universidad Diego Portales, Av Ejercito 441, Santiago,  Chile}
\altaffiltext{2}{Millenium Nucleus ``Protoplanetary Disks in ALMA Early Science", Av Ejercito 441, Santiago, Chile}
\altaffiltext{3}{Departamento de Astronom\'ia, Universidad de Chile, Casilla 36-D, Santiago, Chile}
\altaffiltext{4}{Leiden Observatory, Leiden University, P.O. Box 9513, 2300RA Leiden, The Netherlands}
\altaffiltext{5}{Institute for Astronomy, University of Hawaii at Manoa, Woodlawn Drive, Honolulu, HI, 96822, USA}
\altaffiltext{6}{Department of Astrophysical Sciences, Princeton University, 4 Ivy Lane, Peyton Hall, Princeton, NJ 08544, USA}
\altaffiltext{7}{Departamento de F\'isica y Astronom\'ia, Universidad Valparaiso, Av. Gran Breta\~na
111, Valparaiso, Chile}
\altaffiltext{8}{Harvard-Smithsonian Center for Astrophysics, 60 Garden Street, Cambridge, MA, 
02138, USA.}
\altaffiltext{9}{Joint ALMA Observatory, Alonso de C\'ordova 3107, Vitacura 763-0355, Santiago,  Chile}
\altaffiltext{10}{Australian National University, Mount Stromlo Observatory, Canberra, ACT 2611, Australia}

\textbf{
A snow-line is the region of a protoplanetary disk at which a major volatile, such as water or carbon monoxide, reaches its condensation temperature. Snow-lines play a crucial role in disk evolution by promoting the rapid growth of ice-covered grains$^{1-6}$. Signatures of the carbon monoxide snow-line (at temperatures of around 20 kelvin) have recently been imaged with in the disks surrounding the pre-main-sequence stars TW Hydra$^{7-9}$ and HD163296$^{[3,10]}$, at distances of about 30 astronomical units (au) from the star. But the water snow-line of a protoplanetary disk (at temperatures of more than 100 kelvin) has not hitherto been seen, as it generally lies very close to the star (less than 5 au away for solar-type stars$^{11}$). Water-ice is important because it regulates the efficiency of dust and planetesimal coagulation$^{5}$, and the formation of comets, ice giants and the cores of gas giants$^{12}$. Here we report ALMA images at 0.03-arcsec resolution (12 au) of the protoplanetary disk around V883 Ori, a protostar of 1.3 solar masses that is undergoing an outburst in luminosity arising from a temporary increase in the accretion rate$^{13}$. We find an intensity break corresponding to an abrupt change in the optical depth at about 42 au, where the elevated disk temperature approaches the condensation point of water, from which we conclude that the outburst has moved the water snow-line. The spectral behaviour across the snow-line confirms recent model predictions$^{14}$: dust fragmentation and the inhibition of grain growth at higher temperatures results in soaring grain number densities and optical depths. As most planetary systems are expected to experience outbursts caused by accretion during their formation$^{[15,16]}$ our results imply that highly dynamical water snow-lines must be considered when developing models of disk evolution and planet formation.
}

\begin{figure}
\includegraphics[angle=0, width=17cm, trim = 0mm 20mm 0mm 25mm, clip]{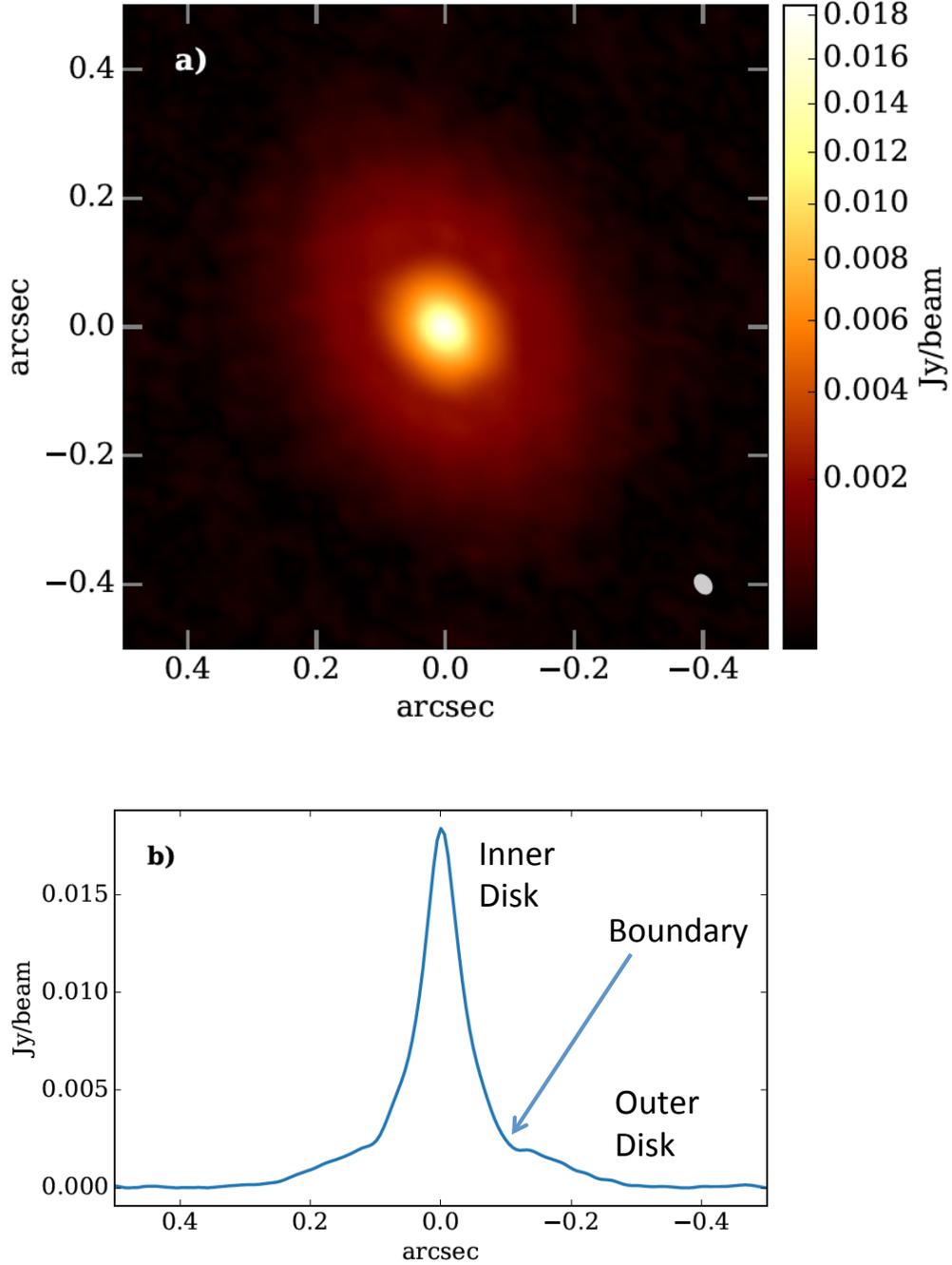}
\caption{ \small
\textbf{ALMA observations of V883 Ori.}
\textbf{(a)} The band-6 image at 0.03$''$ (12 au) resolution obtained on October 27th, 2015. \textbf{(b)} There is a very bright inner disk with radius $\sim$0.1$"$ (42 au), surrounded by a much more tenuous outer disk extending out to radius $\sim$0.3$"$ (125 au). The boundary between these two regions is sharp and probably unresolved. X and Y are the right ascension and the declination, respectively.}
\label{fig:C3_cont}
\end{figure}

V883 Ori is an FU Ori object identified as such by [17] 
from followup spectroscopy of deeply embedded sources from the Infrared Astronomical Satellite (IRAS). 
It is located in the Orion Nebula Cluster, which has a distance of 414$\pm$7 pc$^{[18]}$. It has  a disk mass of  $\gtrsim$0.3 M$_{\odot}$ and a  bolometric luminosity of 400 L$_{\odot} ^{[19]}$. We have obtained 230 GHz/1.3 mm (band-6) observations of V883 Ori using the Atacama Large Millimeter/submillimeter Array (ALMA) in four different array configurations with baselines ranging from 14 m to 12.6 km, which were taken in ALMA Cycle-2 and Cycle-3. These new ALMA observations include continuum and the  $^{12}$CO, $^{13}$CO, and C$^{18}$O  J = 2 --1 spectral lines. We use the C$^{18}$O gas line to investigate the dynamics of the system at 0.2$''$ (90 au) resolution and
the continuum data to constrain the physical properties of the dust in the V883 Ori disk at 0.03$''$ (12 au) resolution.  In Figure~\ref{fig:C3_cont} (top panel) we show our Cycle-3 continuum image at 0.03$''$ resolution, the highest resolution ever obtained for a FU Ori object at millimeter wavelengths.  We find that the V883 Ori disk has a two-region morphology, with a very bright inner disk (r $\sim$ 0.1$''$, 42 au) and a much more tenuous outer disk extending out to $\sim$0.3$''$ (125 au).    The brightness profile (Figure~\ref{fig:C3_cont}, bottom panel) indicates that what looks like a ring at 0.1$''$ in the continuum image is really  a sharp transition between these two regions.
 
\begin{figure}
\includegraphics[width=16cm, trim = 0mm 75mm 0mm 17mm, clip]{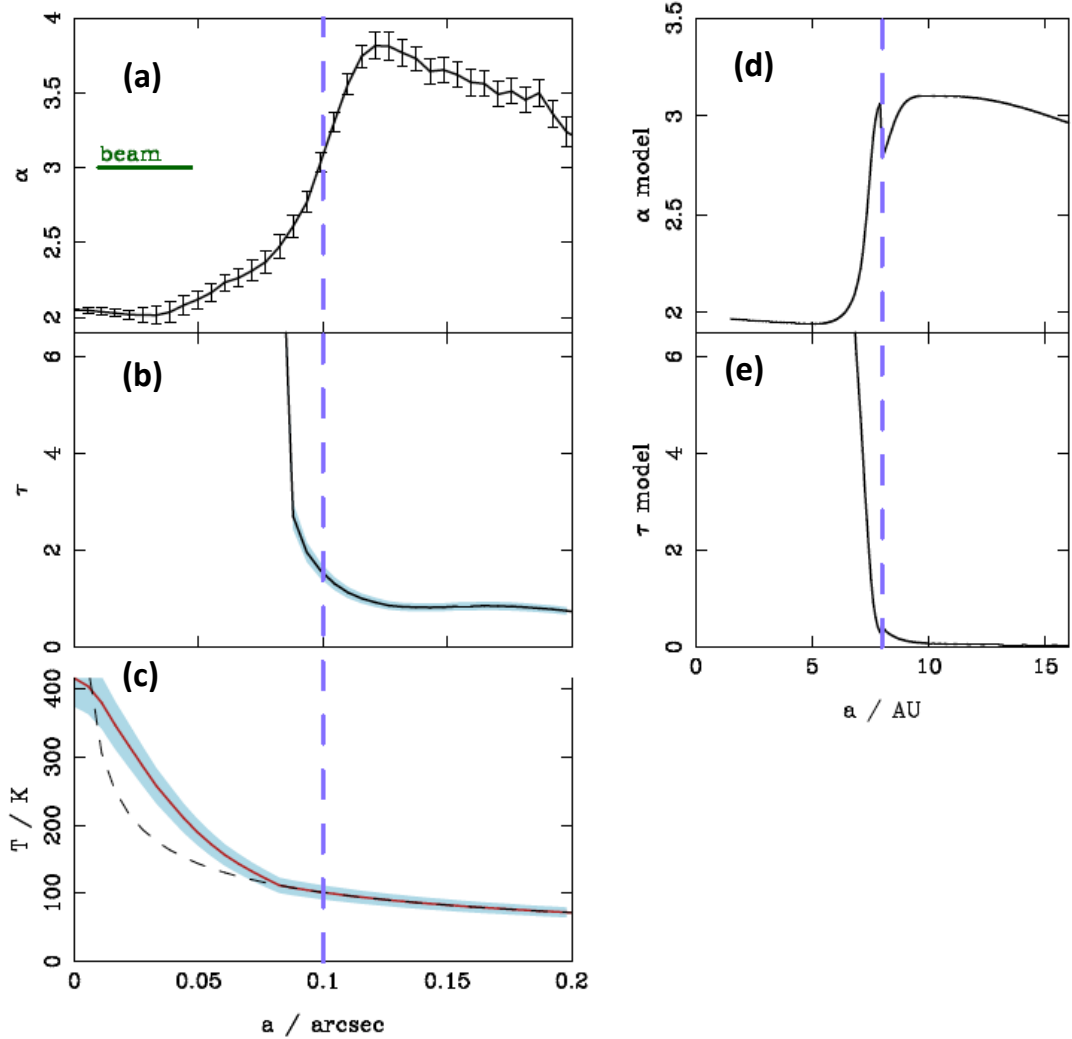}
\caption{ \small 
\textbf{Comparison of observations to models}.
 \textbf{(a-c)},  The spectral index ($\alpha$), the optical depth ($\tau$), and the temperature (T) that we derived from the observed data, all as a function of the semi-major axis, $a$. The temperature profile, shown in red in \textbf{c}, is fixed to a square-root law (shown by the black dashed line) at the radii at which $\tau$ drops below 3. The uncertainties (error bars and light blue regions) are 68\% confidence intervals (see Methods). The blue dashed line corresponds to the location of the water snow-line.  $\alpha$ and $\tau$ are dimensionless. \textbf{d, e,} Predictions from the model in ref. 14 for a disk viscosity of 10$^{-4}$ and an optically thick inner disk.
}
\label{fig:GBD_2}
\end{figure}

Our continuum observations include two different spectral windows  each 1.875 GHz wide and centered at 218.0 GHz and 232.6 GHz, respectively (see Methods section for more details). 
Even though these spectral windows are separated by only 14.6 GHz, the very high signal to noise of our observations allow us to derive accurate information 
regarding the spectral behavior of the spatially resolved disk emission out to r $\sim$0.2$''$ (85 au).  
We use concentric ellipses ($i$ = 38.3 deg and $PA$ = 32.4 deg)  to extract radial profiles in each spectral window  as a function of semi-major axis $a$.
We compare the radial profiles extracted in the two spectral windows separately.
In order to have perfectly matched $uv$-coverage in the two spectral windows, we degraded the  resolution of the 232.6 GHz observations to match that of the  218.0 GHz data.
The spectral index $\alpha$  = ln(F$_{232.6 GHz}$/F$_{218.0 GHz}$) / ln(218.0 GHz/232.6 GHz) as a function of $a$ is  shown in Figure~\ref{fig:GBD_2} (panel \textbf{a}).
We find  distinct spectral behaviors across the disk, with $\alpha = 2.02\pm0.03$ in the central beam of the inner disk, corresponding to optically thick black body emission with T $>$ 100 K, 
and  $\alpha$ reaching 3.7$\pm 0.2$ in the outer disk (typical of optically thin interstellar-medium values).

The observed spectral trends can be cast in terms of physical conditions with grey-body fits$^{20}$ 
that can be used as diagnostics for the optical depth $\tau (a)$ = $\tau_0 \times (\nu/\nu_0)^{\beta}$ and the average dust temperature along the line of sight (summed  up to $\tau \sim 1$).
In our case, the  spectral information available is an amplitude and a slope, at the reference frequency $\nu_0$ = 218.0 GHz.  
Since we are provided with only two data points, 
we fix $\beta$ = 1.0,  as appropriate for circumstellar material$^{21}$.
%
The inner disk is very optically thick, and we can obtain an accurate estimate for T$_s$ that is independent of the adopted $\beta$. 
On the other hand,  $\tau_0$ and $T_s$ become  degenerate in the optically thin regime. 
We therefore adopt a temperature profile with $T_s \propto 1/ \sqrt{a}$ extrapolated from the region where $\tau$=3.
The corresponding $\tau$ and $T_s$ profiles for these assumptions are shown in Figure~\ref{fig:GBD_2} (panels \textbf{b} and \textbf{c}). 
We find that the sharp (unresolved) break at $\sim$0.1$"$ seen in the the V883 Ori disk (see Figure~\ref{fig:C3_cont}, panel \textbf{a}) 
is associated with a steep drop in optical depth and the transition from the optically thick to the optically thin regime.
 This result is robust and insensitive to $\beta$ and to the exact prescription used to estimate  $\tau$ and $T_s$ beyond 0.1$"$. 

This  intensity break occurs where the temperature has dropped below 105 $\pm$ 11 K. 
This temperature is more consistent with water than with the snow-line of any of the other major volatiles in protoplanetary disk (CO, CO$_2$, CH$_4$).
The sublimation temperature of water is a strong function of ambient pressure. While it can be close to $\sim$100~K in the interestellar medium,  high-vacuum laboratory experiments and simulations$^{22-24}$  suggest it should be  $\sim$150-170  K at the 10$^{-4}$ bar pressures expected at the  location of the water snow-line in a typical  disk (1--5 au)$^{[11, 25]}$. However,  since the pressure is lower at  $\sim$40 AU,  the sublimation temperature should also be lower in  the case of  V883 Ori. Furthermore, our temperature estimate is 
based on an extrapolation from the surface of the optically thick inner disk and might underestimate the true temperature of the water snow line due to the intense viscous heating at the disk midplane$^{25}$.   

The observed spectral behavior across the water snow-line has recently been predicted by [14] from numerical models that include radial drift, coagulation, and fragmentation of dust grains.   In Figure~\ref{fig:GBD_2} we show predictions for their models with low disk viscosity ($\alpha_{vis} = 10^{-4}$),  which result in an optically thick inner disk,  as appropriate for V883 Ori.  
The model predictions are not convolved with the ALMA beam and thus have higher resolution than our observations.
In these models, the  fragmentation velocity of dust is 1 m s$^{-1}$ inside the snow-line and 10 m s$^{-1}$ for ice-covered  grains outward of this line.
In this scenario, icy grains quickly grow into cm-size pebbles.  Some of these icy particles drift into the inner disk, where their icy mantles evaporate. When this happens,  their drift velocity decreases, while the fragmentation efficiency increases. 
This produces an accumulation of  mm-size grains in the inner disk, driving the the 230 GHz opacity up and the spectral index to the optically thick limit of $\sim$2. 
In their models,  $\alpha$ increases and $\tau$ decreases steeply around the water snow-line, in remarkably good agreement with our observational results. 
Our ALMA observations thus represent both a confirmation of the predictions by [14] and
the first resolved image of the signature of the water snow-line in a protoplanetary disk.

\begin{figure}
\centering
\includegraphics[width=12cm,  trim = 0mm 00mm 00mm 00mm, clip]{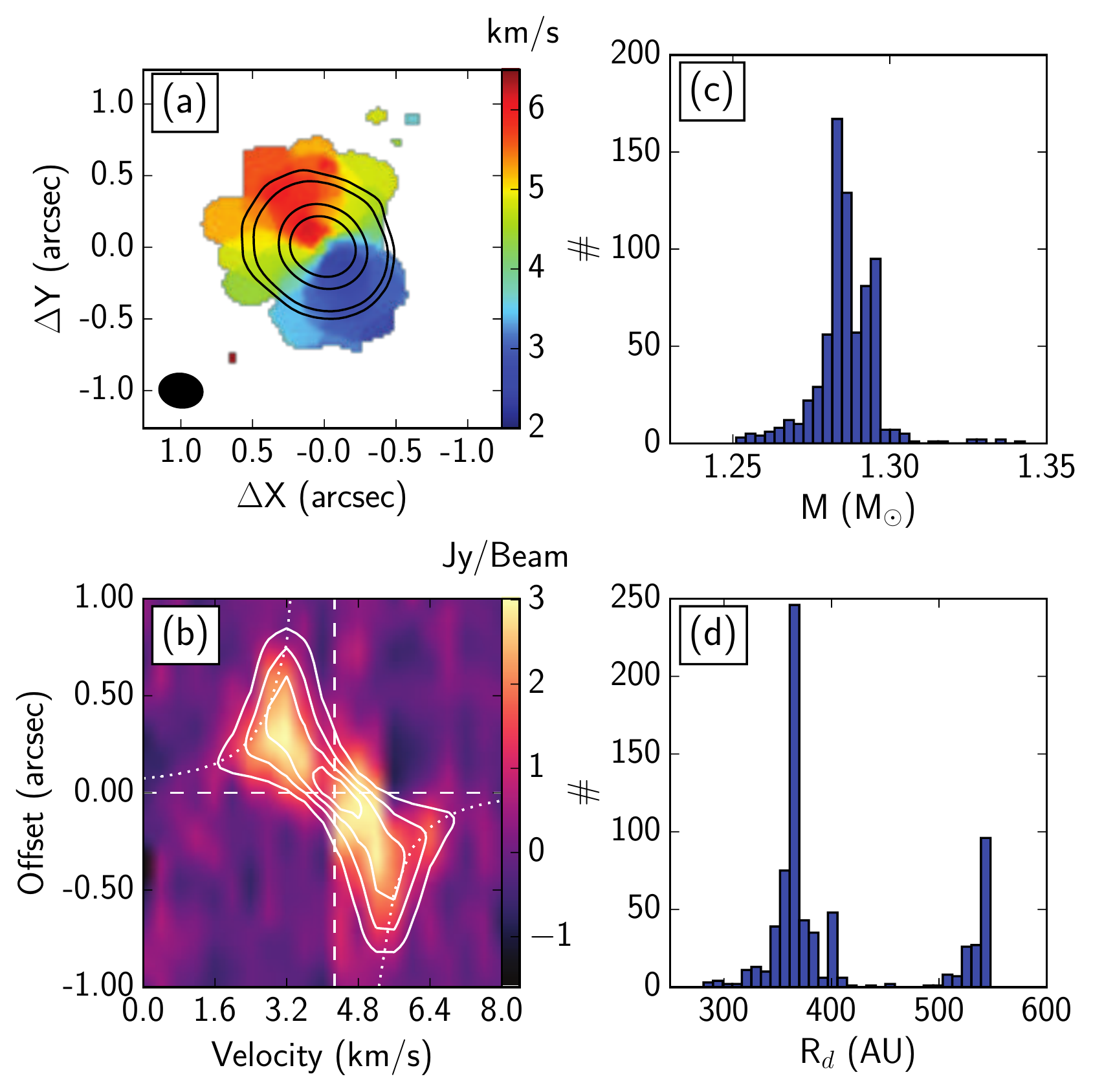}
  \caption{  \textbf{Dynamical mass estimate. a,} Intensity-weighted mean velocity map of the C$^{18}$O line, revealing clear Keplerian rotation. Continuum flux contours have been superposed. X and Y are the right ascension and the declination, respectively. \textbf{b,} Position-velocity diagram of the C$^{18}$O J = 2-1 line. The horizontal and vertical dashed lines denote the position of the source and the source velocity. The curved dotted lines correspond to a Keplerian rotation curve, assuming a central mass of 1.3 M$_{\odot}$. \textbf{c, d,} Distribution of \textbf{c}, the masses of the central object, and
\textbf{d}, the radii of the disk (R$_{d}$), resulting from the minimum $\chi^{2}$ fittings.}
     \label{PVDiagramData}
\end{figure}

By fitting a Keplerian model to the C$^{18}$O line data, we derive a dynamical mass of 1.3$\pm$0.1 M$_{\odot}$ for the central source  (see Figure~\ref{PVDiagramData} and Methods section). 
Adopting an age of 0.5 Myr, as appropriate for a Class I protostar such as V883 Ori$^{[15]}$, its photospheric luminosity should be a mere $\sim$6 L$_{\odot}$$^{[26]}$. 
Based on the stellar mass and  the observed luminosity of 400 L$_{\odot}$$^{[19]}$,   
we derive an accretion rate of  7$\times$10$^{-5}$ M$_{\odot}$yr$^{-1}$,  which is typical of FU Ori objects$^{13}$. 
The location of the water snow-line in a protoplanetary disk is mostly determined by accretion heating in young solar-type stars$^{[11, 25]}$. 
For a 1 M$_{\odot}$ star, the snow-line begins at $\sim$5 au at disk formation and moves inward to $\sim$ 1 au by an age of a few Myr,  driven by the
steady decrease in the accretion rate during disk evolution$^{11}$. 
However, as shown by V883 Ori, this steady evolution is punctuated by extreme bursts of accretion that can drive the snow-line out to $>$ 40 au.

In contrast to the HL Tau protoplanetary disk$^{27}$, 
whose concentric gaps have been interpreted as planet formation occurring at condensation fronts$^{1}$,
the optical depth structure in V883 Ori is close to a step-function, as expected for efficient grain growth beyond a critical radius.
Outward of the water snow-line, grains are covered by ice and can coagulate more efficiently into snowballs and eventually icy planetesimals$^{28}$.
Inside the snow-line, on the other hand, ice mantles evaporate increasing the efficiency of destructive collisions and resulting in the production of a new population of small dust  grains$^{29}$. 
In this scenario, illustrated in the Extended Data Figure 1, an FU Ori outburst  can increase the optical depth at millimeter wavelengths of a large region of the disk by melting snowballs and releasing silicate grains from their icy mantles, which in turn triggers  further dust production. 
If the HL Tau ring system is in fact due to planet formation promoted by the condensation fronts, then the case of
V883 Ori would represent an even earlier stage of disk evolution.  Significant evolution  of solids (growth, migration, and fragmentation) has already occurred, 
but dynamical clearing of gaps by a planet has not yet happened.
While the fact that V883 Ori might show some of the very early steps toward planet formation is fascinating by itself, the implications of an  outward moving snow-line during FU Ori outburst has far-reaching consequences for disk evolution and planet formation in general.   
The water snow-line establishes the basic architecture of planetary systems like our own: rocky planets formed inward of this line in the
protosolar nebula, while giants formed outside.
However, the intimate relation between the position of these snow-lines and the evolution of the central star
is not yet understood. 
While current population synthesis models for planets  do consider a steady decrease in the accretion rate during disk evolution and the corresponding inward motion of the water snow-line at the planet-formation epoch$^{11,25}$ , they do not take into account the dramatic effects FU Ori outbursts have on the snow-line location during the Class I stage.
If most systems experience FU Ori outbursts during their evolution, as proposed by the episodic accretion scenario$^{[13,15,16 ]}$,  this implies that highly dynamical snow-lines must be taken into consideration
by planet formation models. 

\noindent \textbf{References}

\noindent [1]  Zhang, K.; Blake, G. A.; Bergin, E. A. Evidence of Fast Pebble Growth Near Condensation Fronts in the HL Tau Protoplanetary Disk. Astrophys. J. Lett., \textbf{806}, L7 -L12 (2015) \\
\noindent [2]  Okuzumi, S.; Tanaka, H.; Kobayashi, H.; Wada, K.  Rapid Coagulation of Porous Dust Aggregates outside the Snow Line: A Pathway to Successful Icy Planetesimal Formation, Astrophys. J.  \textbf{752},   106 - 123  (2012) \\
\noindent [3] Guidi, G. et al. Dust properties across the CO snowline in the HD 163296 disk from ALMA and VLA observations. Astr. Astrophys. \textbf{588}, 112-123 (2016) \\
\noindent [4] BailliŽ, K.; Charnoz, S.; Pantin, E. Time evolution of snow regions and planet traps in an evolving protoplanetary disk.  Astr. Astrophys. \textbf{577},  65-76 (2015) \\
\noindent [5] Blum, J. $\&$ Wurm, G. The Growth Mechanisms of Macroscopic Bodies in Protoplanetary Disks. Ann. Rev. Astron. Astrophys., \textbf{46}, 21-56 (2008). \\
\noindent [6] Zhang, K. et al. On the Commonality of 10-30 AU Sized Axisymmetric Dust Structures in Protoplanetary Disks,  Astrophys. J. Lett., \textbf{818}, L16 -L22 (2016) \\
\noindent [7] Qi, C. et al. Imaging of the CO Snow Line in a Solar Nebula Analog. Science, \textbf{341}, 630-632 (2013) \\
\noindent [8] Nomura, H. et al.  ALMA Observations of a Gap and a Ring in the Protoplanetary Disk around TW Hya,  Astrophys. J. Lett., \textbf{819}, L7 -L13 (2016) \\
\noindent [9] Schwarz, K. et al. The Radial Distribution of H2 and CO in TW Hya as Revealed by Resolved ALMA Observations of CO Isotopologues,  Astrophys. J, in press.  \\
\noindent [10] Qi, C. et al. Chemical Imaging of the CO Snow Line in the HD 163296 Disk, Astrophys. J.  \textbf{813}, 128-126 (2015) \\
\noindent [11] Kennedy, G. \& Kenyon,  S.   Planet Formation around Stars of Various Masses: The Snow Line and the Frequency of Giant Planets.  Astrophys. J. \textbf{673}, 502-512  (2008) \\
\noindent [12] Morbidelli, A.; Lambrechts, M.; Jacobson, S.; Bitsch, B. The great dichotomy of the Solar System: Small terrestrial embryos and massive giant planet cores. Icarus \textbf{258}, 418-429 (2015) \\
\noindent [13] Audard, M. et al. Episodic Accretion in Young Stars. \textit{Protostars and Planets VI}, University of Arizona Press, Tucson, USA. 387-410 (2014) \\
\noindent [14] Banzatti A. et al. Direct Imaging of the Water Snow Line at the Time of Planet Formation using Two ALMA Continuum Bands. Astrophys. J. Lett.,  \textbf{815}, L15-L20 (2015) \\
\noindent [15] Evans, N. et al. The Spitzer c2d Legacy Results: Star-Formation Rates and E ciencies; Evolution and Lifetimes. Astrophys. J. Suppl. \textbf{181}, 321-350 (2009) \\
\noindent [16] Dunham, M. $\&$ Vorobyov, E. Resolving the Luminosity Problem in Low-mass Star Formation. Astrophys. J.  \textbf{747}, 52-72 (2012) \\
\noindent [17] Strom, K. $\&$ Strom, S.  The discovery of two FU Orionis objects in L1641.   Astrophys. J. Lett., \textbf{412}, L16-L20   (1993) \\
\noindent [18]  Menten, K. M.; Reid, M. J.; Forbrich, J.; Brunthaler, A. The distance to the Orion Nebula.  Astr. Astrophys.   \textbf{474}, 515-520  (2007) \\
\noindent [19] Sandell, G. $\&$ Weintraub, D. On the Similarity of FU Orionis Stars to Class I Protostars: Evidence from the Submillimeter. Astrophys. J. Suppl.  \textbf{134}, 115-132 (2001) \\
\noindent [20] Casassus, S.  et al.   A Compact Concentration of Large Grains in the HD 142527 Protoplanetary Dust Trap.  Astrophys. J.  \textbf{812},  126-139 (2015) \\
\noindent [21] Williams, J. $\&$ Cieza, L.  Protoplanetary Disks and Their Evolutio.  Ann. Rev. Astron. Astrophys.,  \textbf{49},  67-117  (2011)  \\
\noindent [22] Collings, M  et al.   A laboratory survey of the thermal desorption of astrophysically relevant molecules. Mon. Not. R. astr. Soc.  \textbf{354}, 1133-1140   (2004) \\
\noindent [23] Fayolle, E.  et al.  Laboratory H2O:CO2 ice desorption data: entrapment dependencies and its parameterization with an extended three-phase model.   Astr. Astrophys.   \textbf{529}, 74-84     (2011) \\
\noindent [24] Mart\'in-Dom\'enech R.; Mu\~noz Caro, G. M.; Bueno, J.; Goesmann, F. Thermal desorption of circumstellar and cometary ice analogs. Astr. Astrophys.   \textbf{564}, 8-19 (2014) \\
\noindent[25]  Mulders, G.; Ciesla, F. ; Min, M.; Pascucci, I.  The Snow Line in Viscous Disks around Low-mass Stars: Implications for Water Delivery to Terrestrial Planets in the Habitable Zone. 
Astrophys. J.  \textbf{807},  9-15 (2015) \\
\noindent[26]  Siess, L.; Dufour, E.; Forestini, M.  An internet server for pre-main sequence tracks of low- and intermediate-mass stars. Astr. Astrophys.   \textbf{358}, 593-599  (2000) \\
\noindent [27] ALMA Partnership et al.   The 2014 ALMA Long Baseline Campaign: First Results from High Angular Resolution Observations toward the HL Tau Region. Astrophys. J. Lett., \textbf{808}, L13-L12  (2015) \\
\noindent [28] Ros, K $\&$ Johansen, A.  	Ice condensation as a planet formation mechanism.  Astr. Astrophys.   \textbf{523},  137-150 (2013) \\
\noindent [29] Birnstiel, T.; Dullemond, C. P.; Brauer, F.   Gas- and dust evolution in protoplanetary disks.   Astr. Astrophys.   \textbf{513},  79-99 (2010) \\

\noindent \textbf{Acknowledgments:} 

We thank the referees for their valuable comments. We also thank A. Banzatti and P. Pinilla for providing their model predictions in tabular form (Fig. 2d, e). ALMA is a partnership of ESO (representing its member states), NSF (USA) and NINS (Japan), together with NRC (Canada)  and NSC and ASIAA (Taiwan), in cooperation with the Republic of Chile. The Joint ALMA Observatory is operated by ESO, AUI/NRAO and NAOJ. The National Radio Astronomy Observatory is a facility of the National
Science Foundation operated under cooperative agreement by Associated Universities.  Support for this work was provided by the Millennium Science Initiative (Chilean Ministry of Economy), through grant ÒNucleus RC130007Ó and IC120009. L.A.C., D.A.P., J.L.P. and C.C. acknowledge support from CONICYT FONDECYT grants 1140109, 3150550, 1151445 and 3140592, respectively.  H.C. acknowledges support from the Spanish Ministerio de Economia y Competitividad under grant AYA2014-55840P. Our work made use of ALMA data available at https://almascience.eso.org/alma-data with the following accession numbers: 2013.1.00710.S and 2015.1.00350.S.

\noindent \textbf{Author Contributions:} L.A.C. led the ALMA Cycle-2 and Cycle-3 proposals (with the contribution of most co-authors)  and the writing of the manuscript.  S.C. analyzed the Cycle-3 data and performed the grey-body analysis. J.T. and S.B.  determined the stellar dynamical mass.  J.P.W. analyzed the Cycle-2 molecular line data.   S.P. and Z.Z. performed the simulations supporting the Cycle-3 proposal.  All co-authors commented on the manuscript and contributed to the interpretation of the results.

\noindent \textbf{Author Information:} 
This paper makes use of the following ALMA data:   ADS/JAO.ALMA \#2013.1.00710.S and  ADS/JAO.ALMA \#2015.1.00350.S. Reprints and permissions information is available at www.nature.com/reprints. The authors declare no competing financial interests. Readers are welcome to comment on the online version of the paper. Correspondence and requests for materials should be addressed to L.A.C. (lucas.cieza@mail.udp.cl).

\newpage

\newpage

\setcounter{figure}{0}

\makeatletter
\renewcommand{\thefigure}{\empty}
\makeatother

\clearpage

\textbf{METHODS}\label{obs}

Our band-6 Cycle-2 observations were taken under ALMA program 2013.1.00710.S with three different antenna configurations.  V883 Ori was observed on  
December 12$^{th}$, 2014 and April 5$^{th}$, 2015 using 37 and 39 antennas on the C34-2/1 and C34-1/2 configurations, respectively.  The minimum and maximum baselines of  both configurations are very similar, $\sim$14 m  and $\sim$350 m, respectively.  
The integration time was $\sim$2 min in each epoch.
The target was reobserved on  August 30$^{th}$, 2015, with 35  antennas on the C34-7/6  configuration with baselines  in the 42 m  to 1.5  km  range. 
In this configuration, the integration time was $\sim$3 min.
For all antenna configurations, the ALMA correlator was configured so that three spectral windows with 58.6 MHz bandwidths were centered at 230.5380, 220.3987, and  219.5603 GHz to cover the  $^{12}$CO J = 2-1,   $^{13}$CO J = 2-1, and  C$^{18}$O J = 2-1 transitions, respectively.  
Two additional spectral windows with 1.875 GHz bandwidths were centered at 232.6 and  218.0 GHz  for continuum observations.  
Ganymede and J0423-013 were used as  flux calibrators, while the quasars J0538-4405 and J0607-0834 were observed for bandpass calibration. 
Observations of nearby  phase calibrators (J0541-0541, J0532-0307 and/or J0529-0519)  
were alternated with the science target  to calibrate the time dependence variations of the complex gains. 

V883 Ori was also observed in Cycle-3 under program 2015.1.00350.S on October 27$^{th}$, 2015 with 45 antennas in the C38-8 configuration. This the most  extended array configuration offered in Cycle-3, with baselines ranging from 267 m to 12.6 km.  The total on-source  integration time was 23 minutes.
The correlator setup was identical to that of our Cycle-2 observations.
J0541-0541 and J0529-0519 were used as primary and secondary phase calibrator respectively.  J0423-0120 was observed as bandpass
calibrator, and also as primary flux calibrator.
All the data were calibrated using the Common Astronomy Software Applications package (CASA v4.4.0)$^{[30]}$ by the ALMA observatory. 
The standard calibration  included offline Water Vapor Radiometer (WVR) calibration, system temperature correction, bandpass, phase and amplitude calibrations. 
Continuum images and spectral line datacubes were created from the
pipeline calibrated visibilities using the CLEAN routine and Briggs weighting in the
CASA v4.4.0 software package.
Continuum subtraction was performed  in the visibility domain before imaging the CO lines.
Similarly, CLEANing of the dust continuum was performed after removing channels containing line emission.

All the Cycle-2 observations (3 epochs with 3 different array configurations) were combined together  to produce a single 
C$^{18}$O data cube. 
The rms  in this data cube  is 10 mJy/beam per 0.25 km/s channel, with a beam $\sim$0.35$''$ by $\sim$0.27$''$ in size and a  PA of 89.9 (deg).
The long-baseline Cycle-3 dataset was reduced by itself in a similar fashion as the Cycle-2 observations.  
The continuum data resulted in a  0.029$''$ $\times$ 0.038$''$ beam,  and a  rms of 0.05 mJy/beam, after one iteration of phase-only self-calibration. 
The grey-body diagnostics used as proxy for physical conditions of the dust require comparable $uv$-coverages in both continuum frequencies. 
However, the  difference in frequency in  simultaneous observations implies a corresponding radial shift in the $uv$-coverage.  
We followed two independent approaches to build such comparable maps, and confirmed that the two approaches provide very similar spectral index maps and
physical conditions. We first obtained two restored maps with multi-scale Cotton-Schwabb CLEAN$^{32}$ 
 by splitting-off each spectral window, with the same CLEAN masks. We then degraded the higher frequency with an elliptical Gaussian whose axes correspond to the difference in
quadrature of both clean beams. 
We also followed a second method, based on non-parametric Bayesian image synthesis. 
We fit an image model to the observations at each spectral window, and in the visibility domain, by
minimizing the weighted least-square distance,  
as previously performed in other multi-frequency analyses of ALMA data$^{[20, 32]}$.
Since both approaches provided very similar trends, we adopted
the Bayesian image synthesis, as it potentially allows for slightly finer angular detail, 
and the residual were more homogeneous (i.e. free from structure) across the image than the
CLEAN residuals. 
Following [20],  we performed grey-body fits to the our  218.0 and 232.6 GHz images, such that:

\begin{equation}
I_{\nu}(a) =  B_{\nu} (T_s (a))  [1 -\exp (\tau(a)) ].
\end{equation}

where  the optical depth $\tau (a)$ = $\tau_0 \times (\nu/\nu_0)^{\beta}$ and T$_s$ is the average dust temperature along the line of sight (summed  up to $\tau \sim 1$).
In our case, the  spectral information available is an amplitude and a slope, at the reference frequency $\nu_0$ = 218.0 GHz.  
We fixed $\beta$ = 1.0,  as appropriate for circumstellar material$^{21}$.
We estimated the error bars (68$\%$ confidence intervals) on the spectral indices from the rms scatter of specific intensities 
within each elliptical bin summed in quadrature with the rms  intensity of the image synthesis residuals.
The uncertainties on $\tau$ and temperature profiles (light blue regions in Figure~2, panels b and c ) are given by a systematic flux calibration error of 10$\%$ 
(68$\%$ confidence interval).

Molecular line kinematics around a central object are most commonly analyzed by creating a position-velocity (PV) diagram. 
The C$^{18}$O line signal was too weak to be detected at this high spatial resolution of the Cycle-3 observations; therefore, we use the Cycle-2 data at 0.2$"$ resolution to 
investigate the gas dynamics. 
Figure \ref{PVDiagramData} shows the flux as a function of the velocity and position along the major axis of the disk.   In this  diagram, the position is shown as an offset from the center of the disk, and the flux has been integrated along the width of the cut (1.5$"$ in the semi-major axis direction). 
The  disk position angle (PA) was set to 34.2$\pm 2^{\circ}$, determined  from an elliptical Gaussian fit. 
The diagram shows separated blue and red shifted components, suggesting Keplerian rotation. The radius of the disk visible in the diagram is $0.75" \approx \ 320$ AU. The central part around the source velocity (4.3 km  s$^{-1}$) traces the outer slowly rotating material and is largely resolved out due to the extended emission. 
We also see that the data trace the higher velocities to $1.6$ and $6.6$ km s$^{-1}$.  

To further analyze the C$^{18}$O  line emission and give a mass estimate of the central object, we fitted a geometrically thin disk model, based on the model made by [29].
 Based on the channel maps and the PV diagram,  we decided to model a pure Keplerian disk without any infall. The velocity structure is then given by:

\begin{equation}
    v_{\phi}(r) = \sqrt{\frac{GM}{r}}
\end{equation}
We conducted a $\chi^2$-minimization fitting using the method by [34].  
\begin{equation}
    \chi^2 = \frac{1}{N} \sum\limits_{i = 1}^N \left( \frac{D_i - M_i }{\sigma_{rms}} \right)^2,     
\end{equation}
were $D_i$ is  the data, $M_i$ the model and $\sigma_{rms}$ the observed noise error observed for every velocity channel map.
As will method will not  search through the entire parameter-space, multiple runs were done using different random starting parameters.
The parameters used by the fitting algorithm were: the mass of the central object ($M$), the size of the Keplerian disk ($R_d$), the peak intensity ($I_0$) and the full width at half maximum ($fwhm$) of the Gaussian intensity distribution. The PA was again set to $32.4^{\circ}$, while an inclination of $38.3\pm 1 ^{\circ}$ was determined from the ratio of the major and minor axes of the disk in the continuum image. We adopted a distance of $414$ pc$^{[18]}$ 
 while the source velocity ($v_{src}$) was set to  4.3 km s$^{-1}$ based on the moment-1 map. The center of the disk was fixed to the center of the continuum image.  
Of all these runs,  the best fit (lowest $\chi^2$) gave the following parameters: $M = 1.29 \pm 0.02 $ M$_{\odot}$, $R_d = 361 \pm 27$ AU, $I_0 = 0.18 \pm 0.04$ Jy/Beam and $fwhm = 1.14 \pm 0.11''$. 
The distributions of the parameters $M$ and $R_d$ are shown in Figure \ref{PVDiagramData}. A mass around $1.3$ M$_{\odot}$ is the preferred solution. 
We adopt a 10$\%$ total error (68$\%$ confidence interval) in the dynamical mass to account for the uncertainty in the distance to V883 Ori, which dominates the error budget.   
There is more degeneracy in the size of the Keplerian disk, with values mainly varying between 300 and 550 AU. 
Using the parameters given by the best fit, we overlaid contours of the model over the C$^{18}$O emission as shown in Figure \ref{PVDiagramData}. The shape of the contours follow the data well. We see that model traces higher velocities, which is to be expected as the model do not suffer from noise. A Keplerian rotation curve assuming a 1.3 M$_{\odot}$ central mass is also plotted in the figure.  We see that most of the model and C$^{18}$O emission falls well within this curve.

\noindent \textbf{References}

\noindent [30] McMullin, J. P. et al.  CASA Architecture and Applications. Astronomical Society of the Pacific Conference Series,  \textbf{ 376},  127-130  (2007) \\
\noindent [31] Rau, U. $\&$ Cornwell, T.   A multi-scale multi-frequency deconvolution algorithm for synthesis imaging in radio interferometry.   Astr. Astrophys.   \textbf{532},  A71-A87 (2011) \\
\noindent [32]  Cassasus, S.  et al. Flows of gas through a protoplanetary gap. Nature,  \textbf{493}, 191-194  (2013) \\
\noindent [33]  Maret, S.  Thindisk 1.0: Compute the line emission from a geometrically thin protoplanetary disk, Zenodo (doi: 10.5281/zenodo.13823).  \\
\noindent [34]  Powell, M. J.  An  efficient method for finding the minimum of a function of several variables without calculating derivatives.  Computer J. 7, 155-162 (1964).

\newpage

\begin{figure}
\centering
\includegraphics[width=14cm,  trim = 10mm 90mm 0mm 20mm, clip]{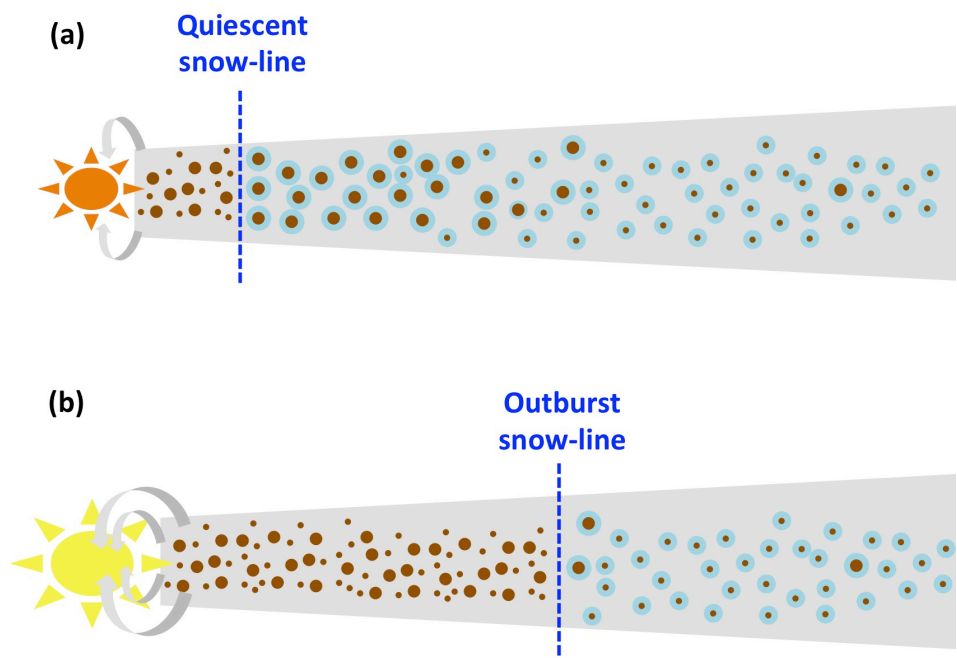}
\caption{
\textbf{Extended Data Figure 1 | Sketch of the observed phenomenon.}. \textbf{a,} During quiescence, the water snow-line around stars of Solar masses is located 5 au or less from the star, where the temperature of the disk reaches the sublimation point of water. \textbf{b,} During protostellar accretion outbursts, this line moves out to more than 40 au, where it can be detected. Outward of the snow-line, grain growth is promoted by the high
coagulation efficiency of ice-covered grains (brown and blue concentric circles). Inward of this line, dust production is promoted by the high fragmentation efficiency of bared silicates (brown circles). This results in the observed break in the disk intensity profile, a steep reduction in the 1.3-mm dust opacity, and a sharp increase in the spectral index across the snow-line.}
     \label{fig:Cartoon}
\end{figure}

\end{document}